\documentstyle[aps,preprint,eqsecnum]{revtex}

\begin{document}

\title{Parallel Algorithm and Dynamic Exponent
for Diffusion-limited Aggregation}
\author{K. Moriarty and J. Machta}
\address{Department of Physics and Astronomy,
University of Massachusetts,
Amherst, MA 01003-3720}
\author{R. Greenlaw}
\address{Department of Computer Science,
University of New Hampshire, 
Durham, NH 03824}

\maketitle

\begin{abstract}
A parallel algorithm for {\em diffusion-limited aggregation\/} (DLA)
is described and analyzed from the perspective of computational
complexity.  The dynamic exponent $z$ of the algorithm is defined with
respect to the probabilistic parallel random-access machine (PRAM) model
of parallel computation according to $T \sim L^{z}$, where $L$ is the
cluster size, $T$ is the running time, and the
algorithm uses a number of processors polynomial in $L$\@.  It is
argued that $z=D-D_2/2$, where $D$ is the fractal dimension and $D_2$
is the second generalized dimension.  Simulations of DLA are carried
out to measure $D_2$ and to test scaling assumptions employed in the
complexity analysis of the parallel algorithm.  It is plausible that the parallel
algorithm attains the minimum possible value of the 
dynamic exponent in which case $z$ characterizes the intrinsic history dependence
of DLA. 
\end{abstract} 

\pacs{61.43.Hv,02.70.-c,05.40.+j}

\narrowtext

\section{Introduction}

This paper examines diffusion-limited aggregation (DLA)~\cite{WiSa,Vi}
from the perspective of computational complexity.  We seek to answer
the following question: Given an idealized parallel computer, what is
the fastest way of generating a representative DLA cluster?  Our
objectives are to give a precise formulation of this question, to
propose a quantitative answer, and to convince the reader that the
answer characterizes an intrinsic property of DLA\@.

A DLA cluster is defined by the following growth process. The cluster
begins as a single, stationary seed particle and grows by the addition
of diffusing particles that stick to the cluster upon contact.  A
diffusing particle (random walker) is released a large distance from
the growing cluster and either joins the cluster by sticking to it or
is discarded if it journeys very far away.  In either case, a new
particle is released as soon as the fate of the preceding one has been
determined. Growth is terminated when a desired cluster mass is
reached.  It is important to note that only one diffusing particle is
present in the system at any given time.  Therefore, it is not obvious
how to take advantage of parallel computation in generating DLA
clusters.

The fractal geometry of DLA aggregates has been extensively
studied~\cite{ToMe,AmMeSt,Os91}.  The clusters bear a strong
resemblance to highly branched structures observed in experiments on
electrodeposition, viscous fingering, crystallization, and the growth
of bacteria colonies~\cite{Vi}.  The asymptotic properties of DLA have
proved difficult~\cite{Vi,Os93} to establish using either theoretical
or numerical methods.  This has created a demand for efficient means
of generating very large aggregates.  To this end, a parallel approach
to DLA has recently been implemented~\cite{KaVeMaWo}.

The speed-up that can be attained by parallelizing a given problem is
the subject of parallel computational complexity theory.  Parallel
complexity theory is the branch of theoretical computer science in
which problems are classified according to the time and processor
requirements of their parallel solutions.  Several growth models
including DLA have been studied from the perspective of parallel
complexity theory.  Eden growth, invasion percolation, ballistic
deposition, and solid-on-solid growth have all been shown to have
highly parallel algorithms~\cite{MaGr}; that is, using sufficiently
many processors (but still polynomial in terms of the system size),
these systems may be simulated in a time that scales as some power of
the logarithm of the system size ({\em polylog time}).  DLA, on the
other hand, has been shown~\cite{Mac93a,MaGr96} to belong to the class
of inherently sequential or, more formally, ${\bf P}$-complete
problems.  Therefore, it is unlikely that DLA clusters can be sampled
in parallel in polylog time when restricted to a number of
processors polynomial in the system size.

Present sequential DLA algorithms~\cite{Os91} achieve running times
that are at best linear in the cluster mass, where cluster mass refers
to the number of particles in the cluster.  Even though the
${\bf P}$-completeness result indicates that a highly parallel (i.e.,
polylog time using a polynomial number of processors) DLA algorithm
probably does not exist, we show that a more modest parallel speed-up
is still possible.  We adopt the conventional theoretical model of
parallel computation known as the parallel random-access machine
(PRAM) and present a polynomial-processor PRAM algorithm for DLA whose
average running time scales as the cluster mass raised to a power less
than unity.

The use of PRAM time permits a robust definition of a {\em dynamic
exponent\/} that can be applied to a wide range of Monte Carlo
algorithms~\cite{MaGr96}.  We define the dynamic exponent $z$ via
\begin{equation}
T \sim L^z ,
\label{eqn:defdynexp}
\end{equation}
where $T$ is the PRAM time needed to generate a representative cluster
of  size $L$ using a number of processors that is bounded by a
power of $L$. The cluster size $L$ 
is the linear dimension of the cluster measured, for example, in
units of the particle size.  The symbol `$\sim$' indicates proportionality in the
asymptotic ($L \rightarrow \infty$) limit.  We will subsequently
determine the value of $z$ for our PRAM algorithm in terms of static
scaling exponents of DLA\@.

Since massive parallelism of the type allowed by the PRAM model is not currently
practical, we do not intend for our DLA algorithm to be used at the present time
for simulations.  Though some elements of our approach may eventually prove
useful in designing a practical parallel algorithm, our primary goal is to
provide an alternative method of characterizing complex objects such as DLA
clusters.  

Bennett~\cite{Benn90,Benn95} suggests that an object should be
regarded as complex if it contains structures that are unlikely to
have arisen quickly.  In this view the presence of unavoidable history
dependence is the signature of physical complexity.  We suggest that
the intrinsic history dependence of a physical object may be
quantified by the PRAM time required to simulate it using the fastest
possible approach.  {\em In this way computational complexity serves
as a gauge of physical complexity.}

The remainder of the paper is organized as follows.
Sec.~\ref{sec:complexity} presents a brief introduction to the theory
of parallel computational complexity.  Sec.~\ref{sec:DLA} places this
paper in context by providing some background on DLA simulation
methods, including a discussion of the parallel approach of
Ref.~\cite{KaVeMaWo}.  Our PRAM algorithm for growing DLA clusters is
presented and discussed in Sec.~\ref{sec:algorithm}, and in
Sec.~\ref{sec:algcomplex} we analyze the algorithm's complexity and
calculate its dynamic exponent.  In Sec.~\ref{sec:sim} we present the
results of a numerical simulation, performed using the sequential DLA
algorithm of Ref.~\cite{Os91}, to test scaling assumptions employed in
Sec.~\ref{sec:algcomplex}.  Sec.~\ref{sec:conclusion} contains our
conclusions.

\section{Introduction to Parallel Complexity Theory}
\label{sec:complexity}

This section provides some background on parallel computational
complexity theory.  The reader is referred to
Refs.~\cite{GrHoRu,GiRy,Papa} for further details.  The objective of
computational complexity theory is to classify problems according to
how the computational resources needed to solve them scale with the
size of the problem.  For parallel complexity the primary resources
are hardware (consisting of memory and processors, or their
equivalents) and time.  One of the strengths of complexity theory is
that resource requirements are comparable within a diverse group of
computational models including parallel random-access machines,
Boolean circuits, and systems of formal logic.  Time requirements for
a wide class of computational models differ by only a logarithmic
factor when the models are required to use polynomially related
amounts of hardware.  Complexity results thus have a rather fundamental
status independent of the computational model adopted.  This fact
supports our belief that a complexity analysis of simulating a
physical system reveals intrinsic properties of the system.

In this paper we employ the parallel random-access machine (PRAM)
model of parallel computation.  A PRAM is composed of a number of
processors, input and output registers, and a global random-access
memory.  The processors are identical except for an identifying
positive-integer label.  Each processor has a local memory and has
access to the common global memory.  The processors run synchronously,
and all execute the same program.  In one time step, a single
instruction is performed by a subset of the processors determined by
the integer labels.  An example of such an instruction is `write the
contents of local memory cell $a$ to global memory cell $b$.'  Note
that $b$ may differ from processor to processor depending on a
previous calculation involving the processor's label. 

It may be that two or more processors will attempt to read from or
write to the same global memory cell during the same clock cycle.  The
way in which such conflicts are resolved distinguishes several
variants of the PRAM model.  These variants all have the same running
time up to logarithmic factors when restricted to using polynomially
related amounts of hardware.  For the sake of this exposition, we
choose the `concurrent read, concurrent write' (CRCW) model.  In the
CRCW PRAM many processors may simultaneously write to the same global
memory cell; of course, a scheme is needed for write arbitration.
There are a number of different methods currently used and we adopt
the one in which the lowest numbered processor writing succeeds.  This
variant of the CRCW PRAM is known as the PRIORITY model and all
references to the PRAM in this paper refer to it.  The word size in a
PRAM is taken to scale as the logarithm of the problem size.

A crucial feature of the PRAM model is that any processor may read
from or write to any global memory cell in a single time step.  Due to
the finiteness of signal speeds and hardware density, PRAM performance
cannot be achieved in a scalable parallel computer.  Nonetheless, the
PRAM model is useful from both practical and theoretical standpoints.
On the practical side, it serves as a guide to the implementation of
algorithms on real parallel machines.  On a conceptual level, PRAM
time provides a measure of a fundamental feature of a computation that
may be called {\em logical depth}~\cite{Benn90}.  Logical depth is the
minimum number of logical operations that must be carried out in
sequence in order to complete a parallel computational process.  The greater the
logical depth, the smaller the speed-up that can be achieved through parallelism.

The power of parallel computation is illustrated by the problem of
adding $n$ numbers.  The problem size in this case is proportional to
$n$ (assuming the numbers are bounded independent of $n$).  On a
sequential random-access machine or familiar desktop computer, $n$
numbers can be added in linear time in an obvious way using a single
DO LOOP\@.  The PRAM approach uses a binary tree.  For simplicity,
suppose that $n$ is an integer power of 2, say, $n$ equals $2^k$.  The
numbers are loaded into global memory, and then each of $n/2$
processors is assigned to add a pair of numbers.  After the first
step, we have $n/2$ partial sums.  These are then added in pairwise
fashion and so on.  Thus after $k$ steps the sum is computed.  The
parallel time is $O(\log n)$ using $n$ processors (we can bring it
down to $n / \log n$ by trading off processors for time) instead of
the $O(n)$ time required by a single processor; so we have achieved an
enormous (exponential) speed-up through parallelism while using a
polynomial number of processors in $n$.  Summing $n$ numbers on a PRAM requires at
least $\log n /\log \log n$ time when restricted to a polynomial number of
processors~\cite{Fi},  so the logical depth of this problem (in terms of the
PRAM) is between $\log n /\log \log n$ and $\log n$.

A similar but somewhat more involved approach may be used to compute
all the partial sums of a list of $n$ numbers in $O(\log n)$ time
using $n / \log n$ processors~\cite{GiRy,AnMi}.  This is an example of
a prefix computation and will be needed later to obtain the full
trajectory of a random walker in an efficient manner.

A problem of size $n$ that can be solved in time $(\log n)^{O(1)}$
({\em polylog\/} time) using $n^{O(1)}$ processors ({\em polynomial\/}
hardware) is said to have a {\em highly parallel\/} solution.
Decision problems (problems with {\sc yes} or {\sc no} answers) that
have highly parallel solutions are in the complexity class ${\bf
NC}$\@.  Eden growth is an example of a model in statistical physics
associated with a decision problem in ${\bf NC}$\@. Eden clusters of
mass $M$ can be simulated on a PRAM in polylog (in $M$) time using
polynomially many processors~\cite{MaGr}.

A problem of size $n$ that can be solved in polynomial ($n^{O(1)}$)
time with polynomially many processors is said to have a {\em
feasible\/} solution.  Decision problems with feasible solutions are
in the complexity class ${\bf P}$\@. (${\bf P}$ is usually defined as
the class of problems that can be solved in polynomial time with a
single processor; however, allowing polynomially many processors does
not enlarge the class since one processor can simulate one clock cycle
of polynomially many processors in polynomial time.)  Clearly ${\bf
NC} \subseteq {\bf P}$\@.  A fundamental question in parallel
complexity theory is whether there are feasible problems that have no
highly parallel solution or, more formally, whether ${\bf NC} \neq
{\bf P}$.

It is conjectured, though not yet proved, that there are in fact
feasible problems that have no highly parallel solution. The best
candidate class of problems so far has been identified using the
property of ${\bf P}$-{\em completeness}.  For a decision problem
$\Pi$ to be ${\bf P}$-complete $\Pi$ must be contained in ${\bf P}$
and {\em all\/} other problems in ${\bf P}$ must be `easily
transformable' into $\Pi$ (see Ref.~\cite{GrHoRu} for further
details).  The ${\bf P}$-complete problems are the hardest problems in
${\bf P}$ to solve in parallel.  It can be proved that if {\em any\/}
${\bf P}$-complete problem has a highly parallel solution then {\em
every\/} problem in ${\bf P}$ has a highly parallel solution.  Thus,
if the conjecture that ${\bf NC} \neq {\bf P}$ holds, ${\bf
P}$-complete problems do not have highly parallel solutions. ${\bf
P}$-complete problems are often referred to as {\em inherently
sequential}.  The conjecture that ${\bf NC} \neq {\bf P}$ is supported
in part by the fact that there is a large class of ${\bf P}$-complete
problems (see Ref.~\cite{GrHoRu}), and, despite much effort, no highly
parallel algorithm has been found for any member of the class. Finding
the shape of a DLA cluster given a list of particle trajectories is a
${\bf P}$-complete problem~\cite{MaGr96}.

While computational complexity theory is generally formulated in terms of decision
problems, computational statistical physics typically deals with {\em sampling
problems}.  The goal here is to generate a representative member of a statistical
ensemble, e.g.~a configuration of Ising spins at a given temperature or a DLA
cluster. Associated with sampling methods in statistical physics are natural
decision problems obtained by considering the random numbers used by the
algorithm as inputs.  Complexity statements concerning sampling methods can be
formulated in terms of these natural decision problems. 
Refs.~\cite{MaGr} and~\cite{MaGr96} discuss the relation between
sampling and decision problems in statistical physics. 

Sampling methods require a supply of random numbers.  Rather than confronting
the subtle issues related to generating random or pseudo-random
numbers, we employ the probabilistic PRAM model in which each processor is
augmented with a device that generates random bits.  In one time step a processor
may draw $w$ random bits, where $w$ is the word size. The algorithm 
described in this paper is a sampling method for DLA implemented on the
probabilistic PRIORITY CRCW PRAM model.

\section{Previous Simulation Methods for DLA}
\label{sec:DLA}

In this section we discuss two simulation approaches for DLA\@.  For
simplicity we restrict the discussion from this point onward to
off-lattice DLA in two dimensions.  We first discuss the standard
sequential method in order to introduce some ideas and terminology
that will be needed for our parallel algorithm. This method will also
be used in the simulations described in Sec.~\ref{sec:sim}.  The
second approach is the parallel DLA (PDLA) method of
Kaufman et al. \cite{KaVeMaWo}.  Their technique is closely related to the
approach that we will use, and its limitations motivate changes that
yield our approach.

The standard sequential simulation method~\cite{ToMe,Os91} implements
several modifications to the original DLA algorithm.  First,
unnecessary initial steps of the walks are eliminated by starting the
walkers at random positions on a `birth circle,' just large enough to
enclose the existing cluster.  This change has no effect on the
cluster distribution sampled by the algorithm.  Efficiency is also
improved, without changing the underlying DLA distribution, by
allowing the walkers to execute variable- rather than fixed-step-size
random walks, taking larger steps in the empty regions away from the
cluster or between its branches.  For our purposes we will assume a
fixed step size.  Finally, if a walker steps outside
of a `death circle,' the walker is discarded and a new one is started
from the birth circle.  If the radius of the death circle is chosen to
be much larger than the cluster radius, deviations from the true DLA
distribution can be made extremely small.  (Issues pertaining to the
birth, death, and step-size of the walkers are discussed in
Ref.~\cite{Vo}.)  A program~\cite{Os91} that employs these techniques
is used in our simulations and achieves a running-time that is very
nearly linear in the cluster mass.

PDLA~\cite{KaVeMaWo} is a practical parallel version of DLA. In this scheme $N$
random walks are controlled by $N$ processors.  As soon as any walker sticks to the
cluster, a new walker is added to the system so that there are always
$N$ diffusing particles.  In the early stages of cluster growth, PDLA
yields more compact structures than ordinary DLA and is similar to
multiparticle diffusive aggregation introduced by Voss~\cite{Voss84}.
Multiparticle diffusive aggregation is not in the same universality
class as DLA\@.  However, as the cluster mass $M$ becomes much larger
than $N$, PDLA crosses over to ordinary DLA.

PDLA becomes a good approximation to DLA for $M \gg N$ for the
following reason.  Consider a group of $N$ walkers launched near a
cluster of mass $M$\@.  We define an {\em interference\/} within such
a group of walkers to be the attachment of one of the walkers to
another member of its group that has already joined the cluster.
Clearly a group of walks performed in parallel may result in a
different cluster configuration than the same group of walks
performed sequentially in some given order.  If $M \gg N$, however, it
is likely that each walker will explore a different region of the
cluster and never have the opportunity to interfere with another
member of its group.  In this case it makes no difference whether the
walks are performed sequentially or in parallel.

Since PDLA uses groups of walks of fixed size $N$, this method has the
same dynamic exponent as the sequential algorithm.  It is only the
prefactor relating running time to cluster mass that is smaller by a
factor of $N$\@.  Our idea is to let the group size be determined by
interferences; i.e., during each iteration, we process the next
interference-free group of walkers in parallel.  Since the average
size of this group will increase with $M$, our algorithm has a smaller
dynamic exponent than PDLA or the sequential algorithm.  Furthermore,
PDLA does not sample the correct DLA distribution except in the limit
$M \gg N$\@.  Our algorithm handles interferences in a way that allows
the correct distribution to be sampled for any value of $M$\@.

\section{New Parallel Algorithm for DLA}
\label{sec:algorithm}

In this section we present our new parallel algorithm for DLA and then
discuss each step of the algorithm in detail.  Our complexity analysis
is somewhat unusual and best described in two sections.  In the
present section we examine the time complexity of each step in terms
of a few parameters involved in the algorithm.  In
Sec.~\ref{sec:algcomplex} we examine the time complexity of the algorithm's main
loop and also explain how several of the parameters are chosen.

The central theme of our algorithm is to generate large and
dynamically increasing groups of non-interfering walkers that in turn
can be processed quickly and correctly in parallel.  At the beginning
of each iteration, we generate, in parallel, a group of random walks
large enough so that an interference will be nearly certain to occur.
Using parallel techniques we then identify the first interference that
would occur if the walks were performed sequentially in a specified
order. Finally, in parallel, we attach any walkers that stick to the
cluster up to the point of the first interference.

The cluster begins as a single seed particle placed at the origin.
The coordinates (pairs of fixed-precision position values) of successive cluster
particles are stored in memory according to the order in which the particles join
the cluster. The algorithm's main loop is iterated until a cluster of
the desired mass is grown.  We analyze the expected number of
iterations of this loop in the next section.  The main loop consists
of the following steps:

\begin{enumerate}

\item 
Choose a birth radius $R_B$, a walk-length $K$, and a number $W$ of
walks to generate.

\item 
Generate $W$ random walks, each beginning at radius $R_B$ and
consisting of $K$ steps of fixed length.  Number the walks from 1 through $W$ to
indicate the order in which they would be performed by a sequential
algorithm.

\item 
Determine the fate of each walker, temporarily ignoring interferences
with the others.

\item 
Identify the first interference that would occur if the walks were
performed sequentially in their specified order.

\item 
Attach any walkers that stick to the cluster up to and including the
second member of the interfering pair identified in Step~4.  Disregard
any remaining walks (note that this does not affect the distribution
of DLA clusters generated). Update the cluster mass $M$ and the
cluster radius $R_C$ accordingly.

\end{enumerate}

We now elaborate on the details of these steps and examine the
time and processor bounds required for each step in terms of several parameters
occurring in the algorithm. The explanation of how these values relate to the
cluster mass $M$ and the analysis of the expected number of iterations of the main
loop are given in the next section.

In Step~1 the radius $R_B$ of the birth circle is chosen, as in most
sequential DLA algorithms, to be a few particle diameters greater than
the distance $R_C$ from the origin to the most remote cluster
particle.  In our algorithm $R_B$ must exceed $R_C$ by at least two
particle diameters to ensure that a single interference cannot cause
the cluster to grow beyond the birth circle. In order to add the
individual steps of a random walk efficiently in parallel, we limit
each walk to a pre-determined number of steps $K$\@.   In principle, walks
in sequential DLA can be arbitrarily long.  Nonetheless, we argue in
Sec.~\ref{sec:algcomplex} that, without affecting the dynamic exponent
of our algorithm, $K$ can be chosen as a function of $R_C$ in such a way that the
ideal DLA distribution is approximated to any desired degree of
accuracy.  In this sense, limiting the walk-length is analogous to
implementing a death circle in sequential DLA\@.  Finally, $W$ is
chosen, as discussed in Sec.~\ref{sec:algcomplex}, to make the
probability of an interference close to unity.  This choice ensures
that the largest possible group of non-interfering walks will be
identified for parallel processing.

To begin a walk in Step~2, a random starting position on the birth
circle is selected.  Then $K$ randomly directed steps of fixed length are
generated in parallel. (Because walks are generated in parallel, the
variable-step-size scheme of the most efficient sequential algorithms cannot be
used.) Finally, a parallel prefix computation~\cite{GiRy} is performed to
calculate the position of the walker after each step of its trajectory.  Since a
prefix computation involving $K$ quantities can be performed on a PRAM in $O(\log
K)$ time using $K/ \log K$ processors, the $W$ $K$-step walks can be determined in
parallel in $O(\log K)$ time using $WK / \log K$ processors.

Once the walks have been computed, we determine in Step~3
if, where, and on what step of its trajectory each of the $W$ walkers
would encounter the existing cluster if none of the other walkers in the group
preceded it.  The following sequence of operations, which determines the fate
of the $i$th walker, is performed for all $W$ walkers in parallel.

First, $M$ processors are assigned to each of the $K$ steps of the $i$th walk, with
the $M$ lowest-numbered processors assigned to the first step and successive
processors assigned to the later steps.  Each of the processors assigned to a
given step of the walk checks one cluster particle to see whether the $i$th walker
would contact it during the specified step.  Any processor that detects such a hit
writes its step number to a memory cell assigned to the $i$th walker.  Note that
all $MK$ processors for the $i$th walk write to the same cell; thus, the
assignment of lower-numbered processors to earlier steps of the walk ensures that
this cell will contain the number of the earliest step (if any) on which the $i$th
walker contacts the cluster in the absence of interference from other walkers.  If
no processor writes to the designated cell, then the $i$th walker does not hit the
existing cluster. The procedure just described can be carried out, for an
arbitrary walker, in constant parallel time using $MK$ processors or,
alternatively, in $O(\log K)$ time using only $MK / \log K$ processors by trading
processors for time.

In the event that the $i$th walker does hit the cluster, its sticking
position and the particle to which it sticks, its `parent' particle,
must be determined.  So far we have identified the step on which the walker hits
the cluster, but, in the process of taking this step, the walker might overlap
several cluster particles.  By means of a standard algorithm for finding the
minimum of $M$ numbers, the first particle contacted during the step (the $i$th
walker's true parent) can be identified in constant parallel time using $M(M-1)/2$
processors or, alternatively, in $O(\log K)$ time using
$M(M-1) / (2\log K)$ processors.  The $i$th walker's position upon first
contacting its parent is recorded as its potential sticking site.

To summarize Step~3, the operations described in the preceding two paragraphs
determine the fate of the $i$th walker as if none of the other walkers in the
group preceded it.  These operations can be performed in constant time and, for
all $W$ walkers in parallel, using $WM\cdot\max\{K,(M-1)/2\}$ processors
or, alternatively, in time $O(\log K)$ using $WM\cdot\max\{K,(M-1)/2\}/ \log K$
processors.

The fourth step of the algorithm is to identify, based on the set of
potential cluster attachments, the next interference
that would occur if the walks were carried out sequentially in the
order specified in Step~2.  In other words we must determine the
number $i_{\text{int}}$ of the lowest-numbered walker which, on some
step of its trajectory prior to striking the existing cluster, would hit a particle
placed at the potential sticking site of some lower-numbered walker.  Determining
$i_{\text{int}}$ can be accomplished in constant time using $KW
(W-1)/2$ processors, one processor for each pair of walks and
each step of the later walker.  Each step of the later walker is
compared to the potential sticking site of the earlier walker.  
If an interference is found during a comparison, then the detecting
processor writes the number of the higher numbered walk of its pair to
the memory location designated $i_{\text{int}}$.  Note that the assignment of the
processors $1, \ldots, K W (W-1)/2$ to their comparisons is such that lower
numbered processors are assigned to comparing lower numbered walks.
For example, processors~1 through $K$ compare walk 1's sticking
site and each step of walk 2, processors~$K+1$ through $3K$ compare
walks 1 and~3, and 2 and~3, and so on.  By again trading processors for time, this
computation can be performed in $O(\log K)$ steps using $KW^2/\log K$ processors.

The techniques from Step~3 can again be used to find the parent and
sticking site of walker $i_{\text{int}}$ taking into account the
addition of lower numbered walks to the cluster.  The necessary
operations can be performed within the bounds noted for an arbitrary walker in
Step~3 above.  Here we observe that no choice of $W$ can guarantee that an
interference will take place.  In the case of no interference, we
simply set $i_{\text{int}}$ equal to $W$.

In Step~5 walker $i_{\text{int}}$ and all lower-numbered walkers that
hit the cluster are permanently placed at their sticking sites.  This
can be accomplished by making a list of the walkers with $i\leq
i_{\text{int}}$ and, by means of a parallel sublist
computation~\cite{GiRy}, removing from the list any walkers that do
not attach to the cluster.  If the initial list is constructed
according to the specified order of the walks, then ranking
~\cite{GiRy} the new cluster particles based on their positions in the
sublist will enable their coordinates to be written to the appropriate
memory locations and the new value of $M$ to be computed.  For an
initial list of length $i_{\text{int}}$, the sublist and ranking
procedures can be carried out in $O(\log i_{\text{int}})$ time using
$i_{\text{int}}/ \log i_{\text{int}}$ processors~\cite{AnMi}.  The
cluster mass may be updated in constant time.  Any change in $R_C$
resulting from the addition of the new particles can be calculated in
constant time using $i_{\text{int}}$ processors since the cluster is
`centered' at the origin.

In the next section, the time and processor requirements found for each step of
the main loop are used to estimate the average running time of the algorithm.

\section{Analysis of the Main Loop of the Algorithm}
\label{sec:algcomplex}

First we describe how $K$ and $W$ are specified, and then examine the expected
number of iterations of the main loop in our parallel DLA algorithm.

To begin the analysis, we must specify how $K$ and $W$ are to be
chosen in Step~1 of the algorithm.  Since $R_B\sim R_C$ and because
random walks behave diffusively, with distance scaling as the square
root of time, choosing $K\sim R_C^2$ is necessary in order to
approximate the ideal DLA distribution. A consequence of this choice
is that the sticking probability remains fixed as the cluster grows.
We note that by increasing the prefactor relating $K$ to $R_C$ we can
come arbitrarily close to sampling the ideal distribution.

The choice of $W$ is not critical as long as the probability of an
interference amongst the walkers remains near unity as the cluster
mass $M$ increases.  Since the sticking probability is constant,
choosing $W \sim M^{1+\epsilon}$ for a small $\epsilon >0$ is sufficient.

For DLA it is believed that the radius of gyration $R_G$ scales with
cluster mass according to
\begin{equation}
R_G \sim M^{1/D},
\label{eqn:massradius}
\end{equation}
where $D$ is the fractal dimension.  Since $R_C\sim R_G$, we have
$K\sim M^{2/D}$; therefore, both $K$ and $W$ need only increase
polynomially with $M$\@.  Consequently, no step of the algorithm
requires more than a polynomial number of processors in $M$\@.
Specifically, the analysis given in
Sec.~\ref{sec:algorithm} yields a processor bound of
$M^{2(1+\frac{1}{D}+\epsilon)}/\log M$ using the probabilistic PRIORITY CRCW PRAM
model.

To estimate the change in cluster mass 
during a single iteration of the main loop, we define $P(M,n)$ to be the
probability for an interference to occur amongst the next $n$ walkers that stick
to a DLA cluster of mass $M$\@.  According to our algorithm for a growing
cluster that has attained a mass $M$, the expected change in cluster
mass during the next iteration is given by
\begin{equation} 
\Delta M = \sum_{n=2}^{W} n \, \left(P(M,n) - P(M,n-1) \right).
\end{equation}
For large $M$ and $W$ the sum can be replaced by an integral,
\begin{equation}
\Delta M \sim \int_{0}^{\infty}dn\,n\,\frac{\partial P(M,n)}{\partial n}.
\label{eqn:DeltaM}
\end{equation}

If DLA clusters are self-similar, then it seems plausible that the
interference probability depends not on $M$ or $n$ individually, but
only on some combination thereof, determined by the multifractal geometry.
Therefore, we make the scaling hypothesis
\begin{equation}
P(M,n) = F(nM^{-\gamma}),
\label{eqn:scalehyp}
\end{equation}
where $\gamma$ is yet to be determined.  Inserting this expression
into Eq.~(\ref{eqn:DeltaM}) yields
\begin{equation}
\Delta M \sim M^\gamma.
\label{eqn:DeltaMgamma}
\end{equation}
In Sec.~\ref{sec:sim} we provide simulation results that support
our scaling hypothesis.

We now determine a theoretical value for $\gamma$ by examining the
form of $P(M,n)$ in the limit $M \gg n \gg 1$.  Clearly, for any
finite $n$, interferences will become rare as $M \rightarrow \infty$.
In this limit, the $n(n-1)/2$ distinct interferences that are possible
amongst the next $n$ walkers that stick to a cluster of mass $M$ may
be treated as independent events.  Thus
\begin{equation}
\lim_{M \rightarrow \infty} P(M,n)= \frac{n(n-1)}{2} P(M,2),
\label{eqn:Plimit}
\end{equation}
where $P(M,2)$ is the probability for an interference to occur between
the next two walkers that stick to a cluster of mass $M$, i.e., the
probability that the second walker will attach to the first.

Now we relate $P(M,2)$ to scaling exponents of the growth probability
distribution.  These scaling exponents, known as generalized
dimensions~\cite{AmCo}, are defined as follows: Cover the accessible
perimeter of a cluster with boxes of linear dimension $l$ and denote
by $\pi_i$ the probability that the next walker to join the cluster
will stick within the $i$th box.  The $q$th moment $Z_q$ of the growth
probability distribution is defined by 
\begin{equation} 
Z_q \equiv \sum_i\pi_i^q.
\label{eqn:Zq}
\end{equation}
For a fixed $l$ such that $a \ll l \ll R_G$ (with $a$ the particle
radius), the scaling behavior of $Z_q$ with $R_G$ defines the $q$th
generalized dimension $D_q$ according to
\begin{equation}
Z_q \sim R_G^{-(q-1)D_q}.
\label{eqn:scaleZq}
\end{equation}

Given two walkers that hit the cluster, we note from
Eq.~(\ref{eqn:Zq}) that $Z_2$ is the probability for both to hit
within the same box.  Clearly $P(M,2)$ is not equal to $Z_2$ since,
for $a \ll l$, only a small fraction of the cases in which two walkers
hit within the same box will result in interferences.  In addition,
there are also cases where two interfering walkers do not hit within
the same box.  Nevertheless, it still seems reasonable to assume that
the scaling behavior of $P(M,2)$ is that of $Z_2$, namely that
\begin{equation}
P(M,2) \sim M^{-\beta},
\label{eqn:P2beta}
\end{equation}
where
\begin{equation}
\beta = D_2/D
\label{eqn:betaD2D}
\end{equation}
and we have made use of Eq.~(\ref{eqn:massradius}).  Simulation
results provided in Sec.~\ref{sec:sim} lend support to Eqs.~(\ref{eqn:P2beta})
and (\ref{eqn:betaD2D}).

Substituting Eq.~(\ref{eqn:P2beta}) into Eq.~(\ref{eqn:Plimit}) gives,
for $n \gg 1$, $P(M,n)\sim n^2M^{-\beta}$.  Thus, if $P(M,n)$ has the
assumed form (\ref{eqn:scalehyp}), the exponent $\gamma$, defined in
Eq.~(\ref{eqn:scalehyp}), is related to the exponent $\beta$, defined
in Eq.~(\ref{eqn:P2beta}), by
\begin{equation}
\gamma=\beta/2,
\label{eqn:gammabeta}
\end{equation}
where $\beta$ has the theoretical value $D_2/D$.

We now consider the time requirements of our algorithm.  Specifically,
we obtain an expression for $\Delta T$, the expected time required to
complete an iteration during which a cluster of mass $M$ grows by an
amount $\Delta M$\@.  In Sec.~\ref{sec:algorithm}, we found that
Steps~1--4 of the algorithm could be performed in $O(\log K)$ time and Step~5 in
$O(\log i_{\text{int}})$ time.  Since $\Delta M$ is simply equal to the
product of the (constant) sticking probability and the expected value
of $i_{\text{int}}$, the time to perform Step~5 scales as $\log \Delta
M$\@.  As previously mentioned, both $K$ and $\Delta M$ scale as
powers of $M$; therefore,
\begin{equation}
\Delta T \sim \log M.
\label{eqn:DeltaT}
\end{equation}

Combining Eqs.~(\ref{eqn:DeltaMgamma}) and (\ref{eqn:DeltaT}) we find
that the growth rate $dM/dT$ ($\approx \Delta M/\Delta T$) of a
typical cluster is given by $dM/dT \sim M^\gamma / \log M$\@.
Integrating this expression yields
\begin{equation}
T \sim M^{1-\gamma}\log M.
\label{eqn:T}
\end{equation}
By Eqs.~(\ref{eqn:massradius}), (\ref{eqn:betaD2D}),
(\ref{eqn:gammabeta}), and (\ref{eqn:T}) it follows (ignoring the
logarithmic factor) that $T \sim R_G^z$, where the dynamic exponent is
\begin{equation}
z=D-D_2/2.
\label{eqn:zd2}
\end{equation}
It should be noted that this expression for $z$ is valid for any
dimension greater than one, and depends on dimension only through the
static scaling exponents $D$ and $D_2$.

\section{Simulation Results}
\label{sec:sim}

In order to test our assumptions, Eqs.~(\ref{eqn:scalehyp}) and
(\ref{eqn:P2beta}), we grew a total of 1440 DLA clusters, using a fast
sequential algorithm developed by Ossadnik~\cite{Os91}.  We measured
the probability $P(M,n)$ for an interference to occur amongst the next
$n$ walkers that stick to a cluster of mass $M$\@.  Each cluster was
assigned a specific mass $M$ and a particular value of $n$ for which
$P(M,n)$ would be measured.  Once a cluster reached mass $M$, repeated
trials were performed to determine the likelihood of an interference
occurring amongst the next $n$ walkers to stick.  We performed
${10}^6$ trials for $n=2$ and ${10}^3$ trials for $n > 2$.  Each trial
consisted of releasing test walkers either until some test walker
attached to a previous test walker (interference) or until $n$ walkers
had successfully attached to the cluster without such an interference
occurring.  In both cases all test walkers were then removed from the
cluster and the next trial was begun.  The fraction of trials for
which an interference occurred provided a measurement of $P(M,n)$ for
the particular cluster being tested.

In this fashion we measured $P(M,n)$ for clusters at each of eight
masses ranging from ${10}^5$ to $1.7 \times {10}^6$ and for each of
nine $n$ values between 2 and 94.  For each $(M,n)$ pair, we grew a
sample of 20 clusters and calculated the mean $\langle P(M,n) \rangle$
of the 20 individual measurements along with its standard error.
Although, at first glance, using each cluster for only one $(M,n)$
pair might seem inefficient (as opposed, say, to making measurements
at intermediate masses during the cluster's growth), our method
eliminates undesirable correlations between data points.  The entire
experiment required about four weeks of CPU time on a DEC
Al\-pha\-Sta\-tion 200 4/166.

To check Eq.~(\ref{eqn:P2beta}) and to measure $\beta$, we fit a
line to a plot of $\log_{10} \langle P(M,2)\rangle$ vs.\ $\log_{10} M$ 
as shown in Fig.~\ref{fig:P2beta}.  Statistical error bars (not shown)
in the vertical direction are approximately equal to the symbol
height.  The apparent good fit indicates that the dependence of
$P(M,2)$ on $M$ is well described by a power law for the range of
masses we considered.  Our best-fit value of the exponent in
Eq.~(\ref{eqn:P2beta}) is $\beta_{\text{expt}}=.530 \pm .003$. The
error bounds in this and subsequent exponent estimates include only
statistical contributions and do not reflect the uncertainty
associated with extrapolating to infinite cluster size.

In order to test Eqs.~(\ref{eqn:scalehyp}) and (\ref{eqn:gammabeta}),
we plotted $\langle P(M,n) \rangle$ vs.\ $nM^{-\beta_{\text{expt}}/2}$ 
as shown in Fig.~\ref{fig:scale}. For the sake of clarity, only three
of the cluster masses are represented in the plot, but the observed
data collapse is no less convincing when all the data are included.
Thus, for our range of cluster masses (a factor of 17), the
interference probability does appear to have the scaling form assumed
in Eq.~(\ref{eqn:scalehyp}) with $\gamma=\beta/2$.

Now we compare our measured value of $\beta$ with the prediction
$\beta = D_2/D$ (Eq.~(\ref{eqn:betaD2D})).  A numerical estimate for
the fractal dimension of two-dimensional DLA is $D = 1.715 \pm
.004$~\cite{ToMe}.  Since $D$ is known much more precisely than $D_2$,
we use this value of $D$ and our measured value of $\beta$ to predict
$D_2$.  We obtain $D_2=.909 \pm .006$ as compared with $D_2=.83 \pm
.05$ in Ref.~\cite{BaSp} and $D_2=.980 \pm .010$ in
Ref.~\cite{HaMePr}.  Since the two previous results are inconsistent
with each other and since our result, though intermediate between the
two, is consistent with neither, the question of whether $\beta =
D_2/D$ remains open.  At present, we see no reason to abandon
Eq.~(\ref{eqn:betaD2D}); therefore, we regard our simulation as
providing a new, relatively precise measurement of $D_2$\@.  Using our
numerical value of $\beta$ along with Eqs.~(\ref{eqn:gammabeta}) and
(\ref{eqn:T}) yields a numerical value $z = 1.261 \pm .004$ for the
dynamic exponent of our parallel DLA algorithm, as compared with $z =
D \approx 1.7$ for both PDLA and the best sequential DLA algorithms.

\section{Conclusions}
\label{sec:conclusion}

The DLA growth process is inherently history dependent. The random
walks that generate the cluster must in principle be run one at a time
to precisely simulate the DLA distribution.  Previous ${\bf
P}$-completeness results show that there is almost certainly no clever
way to fully eliminate this history dependence and to generate DLA
clusters from walk trajectories by any
highly parallel process.  In this paper we have demonstrated that a
more modest parallel speed-up is possible.  We have shown that an average running
time sublinear in the cluster mass may be achieved by processing walkers in
successive, interference-free groups.  
The interference probability determines how quickly the average size
of these groups increases with cluster mass and thus controls the
extent of the speed-up attainable by this approach.

By adopting the PRAM model of parallel computation, we are able to
precisely characterize the speed-up achieved by our parallel approach
to DLA\@.  The dynamic exponent $z$ for the algorithm relates the
average PRAM time $T$ to the cluster mass $M$ via $T \sim M^{z/D}$\@.  
By means of simple scaling assumptions involving the interference
probability, we have argued that the dynamic exponent may be expressed
in terms of static exponents according to $z = D-D_2/2$, where $D$ is
the fractal dimension and $D_2$ is the second generalized dimension.
For two dimensions we find that $T \sim M^{0.74}$, whereas the running time for the
best possible sequential algorithm is at least linear in $M$. 
Though we have not directly tested the parallel algorithm, we have performed
sequential DLA simulations whose results support our scaling assumptions.  In
addition, from measurements of relevant interference probabilities, we have
extracted a new value of $D_2$ that lies between two previously published values.

Since different dynamics may yield the same distribution of
structures, it is possible that an entirely new method will be
discovered to simulate DLA that can be implemented in parallel with a
better speed-up.  The ${\bf P}$-completeness
results~\cite{Mac93a,MaGr96} for the known DLA dynamics do not rule
out the possibility that such a new method exists and perhaps even
runs in polylog time using a feasible number of processors.  As an
example of this, we note that the usual rules for growing Eden
clusters~\cite{Vi} lead to a ${\bf P}$-complete problem.  There is,
however, an entirely different approach to creating Eden
clusters~\cite{MaGr} that can be implemented in polylog time using a
polynomial number of processors on a PRAM\@.  Nevertheless, given the
considerable effort that has gone into understanding and simulating
DLA, we believe it is unlikely that there is an entirely new and
highly parallel method of sampling DLA clusters.  The present evidence
suggests that DLA has qualitatively greater logical depth than Eden
growth and related models.

Interferences between the random walkers seem to provide the
fundamental limitation to parallelizing DLA\@.  Our algorithm works by
processing in parallel, at each stage, the initial maximal group of
non-interfering walkers.  Therefore, it seems that a more
sophisticated method of processing interferences in parallel would need
to be developed if our algorithm is to be improved.  At the present
time we have not been able to devise such a technique.  For the moment
suppose that our algorithm is actually optimal in the sense that no
other algorithm for sampling DLA has a smaller value of $z$.  Because
of the equivalence, up to logarithmic factors in the time, of
differing models of parallel computation, the minimum value of $z$ is
a well-defined quantity that characterizes the DLA distribution.
Assuming we have actually found the fastest PRAM algorithm for DLA, we
have measured the intrinsic history dependence (equivalently, logical
depth) of DLA\@.  In any case, our algorithm is a new technique for
parallel generation of DLA clusters and provides an upper bound
on the time complexity of producing these clusters.

\section*{Acknowledgements}
We are grateful to Peter Ossadnik for providing us with the DLA code
used in the simulations and to the UMass high energy theory group for
use of their AlphaStation.  We thank David Barrington, Peter Ossadnik,
William Leonard, and Stefan Schwarzer for useful discussions and
correspondence.  This work was supported in part by NSF Grants
DMR-9311580 and DMR-9632898.

\newpage

\begin{figure}
%\centerline{\psfig{figure=fig1.ps}}
\caption{$\log_{10} \langle P(M,2)\rangle$ vs.\ $\log_{10}M$,
where $\langle P(M,2) \rangle$ is the mean probability, from a sample
of 20 DLA clusters of mass $M$, for an interference to occur between
the next two walkers that stick.  The solid line is a linear fit to the
data and has slope $-.53$.}
\label{fig:P2beta}
\end{figure}

\begin{figure}
%\centerline{\psfig{figure=fig2.ps}}
\caption{$\langle P(M,n) \rangle$ vs.\ $nM^{-\beta_{\text{expt}}/2}$
plotted for $M = 1 \times 10^5$ ($\circ$), $5 \times 10^5$ ($\Box$), and
$1.7 \times 10^6$ ($\bigtriangledown$). $\langle P(M,n) \rangle$ is the
mean probability, from a sample of 20 DLA clusters of mass $M$ for an
interference to occur amongst the next $n$ walkers that stick, and
$\beta_{\text{expt}} = .53$ is minus the slope of the line shown in
Fig.~1.}
\label{fig:scale}
\end{figure}

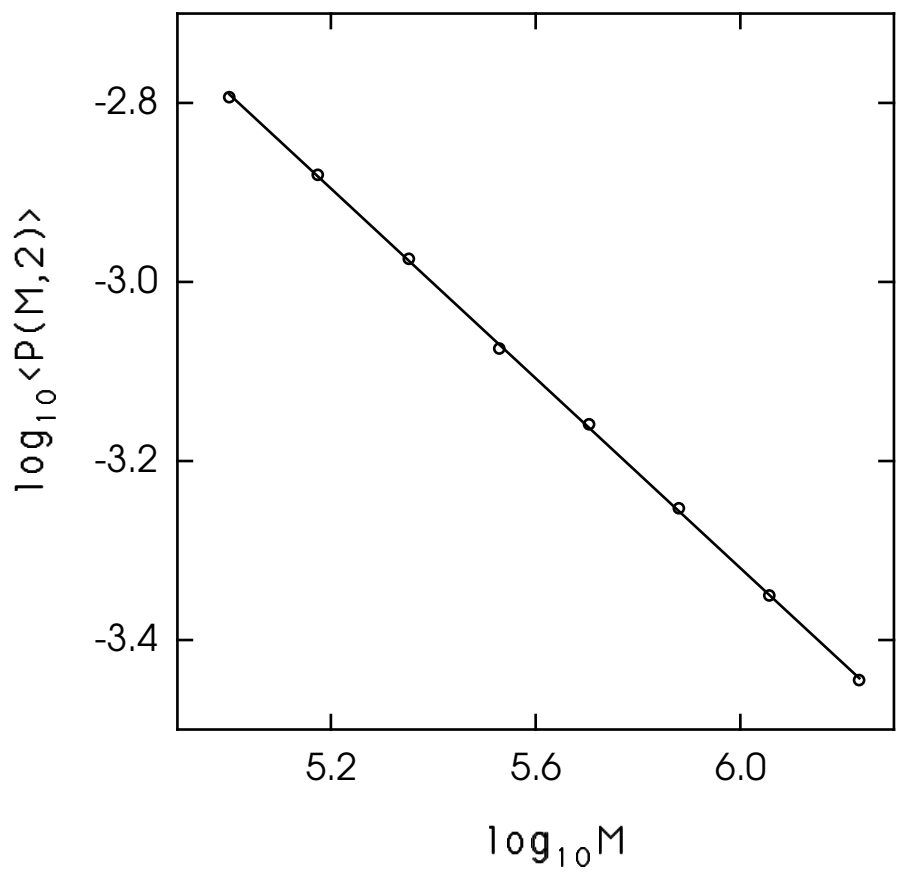
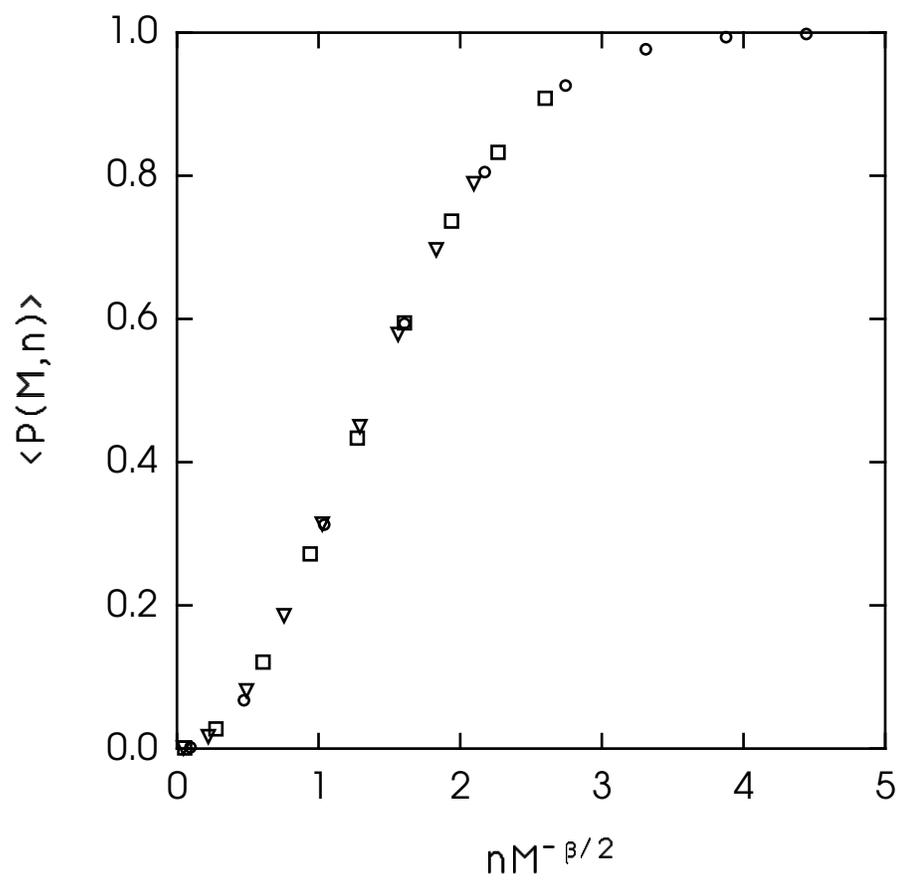
\end{document}